\documentclass[showpacs,twocolumn,prb]{revtex4}
\usepackage{graphicx}
\newcommand{\be}{\begin{equation}}
\newcommand{\ee}{\end{equation}}
\newcommand{\bea}{\begin{eqnarray}}
\newcommand{\eea}{\end{eqnarray}}
\newcommand{\bwt}{\begin{widetext}}
\newcommand{\ewt}{\end{widetext}}
\newcommand{\ra}{\rangle}
\newcommand{\la}{\langle}
\newcommand{\up}{\uparrow}
\newcommand{\dn}{\downarrow}
\newcommand{\ups}{|+\rangle}
\newcommand{\dns}{|-\rangle}

\newcommand{\Gs}{|G\rangle}
\newcommand{\Vs}{|0\rangle}
\newcommand{\Gc}{\langle G|}
\newcommand{\Vc}{\langle 0|}
\begin{document}
\title{Impurity Scattering of Wave Packets on a Lattice}
\author{Wonkee Kim$^1$, L. Covaci$^1$, and F. Marsiglio$^{1-3}$}
\affiliation{ $^1$Department of Physics, University of Alberta,
Edmonton, Alberta, Canada, T6G~2J1 \\
$^2$DPMC, Universit\'e de Gen\`{e}ve, 24 Quai Ernest-Ansermet,
CH-1211 Gen\`{e}ve 4, Switzerland, \\
$^3$National Institute for Nanotechnology, National Research Council
of Canada, Edmonton, Alberta, Canada, T6G~2V4}

\begin{abstract}
Quantum transport in a lattice is distinct from its counterpart in
continuum media. Even a free wave packet travels differently in a
lattice than in the continuum. We describe quantum scattering in a
one dimensional lattice using three different formulations and
illustrate characteristics of quantum transport such as resonant
transmission. We demonstrate the real time propagation of a wave
packet and its phase shift due to impurity configurations. Spin-flip
scattering is also taken into account in a spin chain system. We
show how individual spins in the chain evolve as a result of a
spin-flip interaction between an incoming electron and a spin chain.
%
%
\end{abstract}

\date{\today}
\maketitle

\section{introduction}

In the last few decades the advent of pump-probe optical methods
\cite{drescher,baltuka,hegman03} and
similar measurement techniques has stimulated
interest in time-dependent phenomena in physical systems.\cite{freeman}
For example, questions concerning the details of magnetization reversal
in ferromagnetic thin films can be addressed experimentally
\cite{choi01} and theoretically \cite{rikvold02}. Another example is
polaron formation in a semiconductor, where both experimental and
theoretical work are just starting.\cite{trugman03} While much of
the work on magnetization reversal has utilized a classical or at
most semi-classical description,
\cite{berger96,slonczewski96,slonczewski99,berger01,llg1,llg2}
more recent work has focused on a fully quantum mechanical
description.\cite{kim05} Such a microscopic description is expected
to be necessary and insightful for small (i.e. quantum dot) systems.

Insofar as many of these phenomena occur in the solid state, the
underlying lattice structure may play an important role. Concerning
the theoretical description of magnetization reversal, calculations
to date have either focused on simple models,
\cite{berger96,slonczewski96,slonczewski99,berger01,kim04}
or have tried to utilize
realistic transport equations with band structures relevant to the
materials of interest.\cite{stiles} In this work we wish to solve
simple scattering models based on tight-binding band structures.
While we will consider mainly scalar potential scattering, the
formalism is extendable to spin-flip scattering, which will be
discussed. In particular, an incoming electron can scatter off a
ferromagnetic thin film modelled by a Heisenberg Hamiltonian, and we
can monitor the real time reaction of the magnetization to the
onslaught of electrons with a completely quantum mechanical
description. We also want to utilize a framework that is amenable to
numerical calculation. By this we mean the following: as
interactions are introduced, problems will become formidable by
analytical means and large scale computation will be required. Most
often this means Monte Carlo methods (at least this is so in
equilibrium and linear response calculations so far) which are often
well-suited to simple lattices. For these reasons we believe it is
beneficial to have a lattice-based framework to tackle
non-equilibrium phenomena in solid state systems.

We begin with a description of the non-interacting electron, where
already a novel property emerges due to the lattice: the degree of
spreading of a propagating wave packet can be controlled by
judicious choice of the electron energy. This is always true in one
dimension and has more limited validity in higher dimensions. Next
we solve the scattering problem for a rectangular potential barrier
and other, simple barriers (or wells) that enter into impurity
problems. We outline the methodology to solve the problem
numerically, and use some illustrative examples to demonstrate the
accuracy and efficiency of these calculations. Finally, we discuss
spin flip scattering in a spin chain.

This paper is organized as follows. In Section.~II, we describe wave
packet transport in a lattice without scattering, and a possible
solitonic behavior which is impossible in the continuum limit. In
order to illustrate quantum scattering off impurities in a lattice,
we use a quantum mechanical approach and the transfer matrix
formalism, and compare the two methods using an example in
Section~III. A direct diagonalization of the Hamiltonian will
illustrate the time evolution of a wave packet in a lattice with
embedded impurities in Section~IV. In Section~V, we briefly explain
a procedure to study spin flip scattering on a lattice. In
section~VI we summarize our results and outline possible future
directions.

\section{wave packet transport}

In order to understand the differences between calculations of
quantum mechanical phenomena in a lattice from their counterpart in
the continuum limit, it is appropriate to begin with transport of a
free wave packet. As is normally done in textbooks,\cite{shankar80}
one can introduce a free Gaussian wave packet in the continuum
limit:
\begin{equation}
\Psi(x,0)=\frac{1}{\left(2\pi\alpha^{2}\right)^{1/4}}
e^{-\frac{1}{4}(x-x_{0})^{2}/\alpha^{2}+ik_{0}(x-x_{0})}
\label{free}
\end{equation}
where $x_{0}$ and $k_{0}$ are the mean position and momentum,
respectively, of the wave packet, and $\alpha$ is the position
uncertainty associated with the wave packet. To see the time
evolution of the wave packet we expand $\Psi(x,0)$ in terms of the
momentum eigenstates $|k\ra$. They are the known solutions to the
Schrodinger equation in free space; hence the time dependent
Schrodinger equation is readily solved in this basis. The result is
\cite{shankar80}
\be \Psi(x,t)=\left(\frac{\alpha^{2}}{2\pi}\right)^{1/4}
\frac{e^{i(k_{0}x-E_{0}t)}}{\sqrt{\alpha^{2}+it/2m}}
e^{-\frac{1}{4}(x-v_{0}t)^{2}/(\alpha^{2}+it/2m)}
\label{free_time-evoloved} \ee where $E_{0}=k^{2}_{0}/2m$ and
$v_{0}=k_{0}/m$ are the average energy and particle velocity,
respectively. Since $\la x\ra=v_{0}t$ and $\la
x^{2}\ra=(v_{0}t)^{2}+ \left[\alpha^{2}+(t/2m\alpha)^{2}\right]$,
the position uncertainty $\Delta x$ defined as $(\Delta x)^{2}=\la
x^{2}\ra -\la x\ra^{2}$ becomes $ \Delta
x=\sqrt{\alpha^{2}+t^{2}/(2m\alpha)^{2}} $. Similarly, we obtain the
momentum uncertainty $\Delta k=1/(2\alpha$). The uncertainty
relation is therefore%
\be \Delta x\cdot\Delta k=\frac{1}{2}
\sqrt{1+\left(\frac{t}{2m\alpha^{2}}\right)^{2}}.
\label{uncertainty_time} \ee%
This means that the uncertainty
relation increases as a function of time. Also note that the
relation does not depend on the mean momentum of the wave packet; as
we will see, this is true only for a parabolic energy dispersion.

In a one dimensional lattice described by a nearest neighbor
tight-binding model, $E_{k}=-2t_{0}\cos(ka)$, where $t_{0}$ is a
hopping amplitude to the nearest neighbor site and $a$ is a lattice
constant. Hereafter we set $a=1$ and use it as the unit of length.
The position is now discrete and represented by $x_{i}$, with $i$
the lattice site label. Upon expanding in terms of the momentum
eigenstates in a box with periodic boundary conditions, one obtains
\be \Psi(x_{i},t)=\left(\frac{\alpha^{2}}{2\pi^{3}}\right)^{1/4}
\int^{\pi}_{-\pi}{}dk\;
e^{ikx_{i}-\alpha^{2}(k-k_{0})^{2}-iE_{k}t}\;.
\label{psi_in_latt} \ee Note that the integration is over a
Brillouin zone, due to the discreteness of the
lattice; nonetheless, if $\alpha$ is large enough, the integration
range can be extended from $-\infty$ to $\infty$ without altering
the integral. In the same way, using a large $\alpha$ expansion for
the exponent in Eq.~(\ref{psi_in_latt}), and keeping terms up to
${\cal O}(1/\alpha^{2})$, one can convert the integral into a
Gaussian integral, as in the continuum limit. Performing the
integration, we obtain
\bwt
\be
\Psi(x_{i},t)=\left(\frac{\alpha^{2}}{2\pi^{3}}\right)^{1/4}
\frac{\sqrt{\pi}}{\sqrt{\alpha^{2}+itE_{k_0}^{\prime \prime}/2}}
e^{ik_{0}x_{i}- iE_{k_0}t} e^{-\frac{1}{4}[x_{i}-
tE_{k_0}^{\prime}]^{2}/[\alpha^{2}+itE_{k_0}^{\prime \prime}/2]}\;,
\label{psi_time} \ee
\ewt
where we restored more generality (than in
Eq. \ref{psi_in_latt}) by using $E_{k_0}^\prime$ and
$E_{k_0}^{\prime \prime}$ to refer to the first and second
derivatives of the dispersion, $E_k$, with respect to momentum $k$,
and evaluated at $k_0$. For a quadratic dispersion one readily
obtains the result Eq. \ref{free_time-evoloved}.
On the other hand,
for the nearest
neighbor model, $E_{k_0}^\prime \equiv v_{k_0} = 2t_0 \sin(k_0)$
and $E_{k_0}^{\prime \prime} \equiv v_{k_0}^\prime = 2t_0
\cos(k_0)$, where $v_{k_0}$ is the group velocity, and
$v_{k_0}^\prime$ is the group velocity dispersion.
The expansion is valid as long as $t\ll\left(\alpha/k_{0}\right)^{3}/t_{0}$.
For $k_{0}=\pi/2$, the validity length $l$ is $l=v_{k_{0}}t\sim \alpha^{3}$.
If $\alpha$ is on a nanometer scale, $l$ is of order $1 \mu m$.

We are now able to calculate the uncertainty relation for the
lattice case with nearest neighbor hopping only, at any time $t$:
$\la x \ra = 2t_{0}\sin(k_{0})t$ and $\la x^{2} \ra =
(2t_{0}\sin(k_{0})t)^{2}+\alpha^{2}+
(t_{0}\cos(k_{0})t/\alpha)^{2}$. The uncertainty in position is then
$\Delta x = \sqrt{\alpha^{2}+t^{2}_{0}\cos^{2}(k_{0})t^2/\alpha^{2}}
$. The uncertainty in the momentum is the same as in the continuum
limit; namely, $\Delta k=1/2\alpha$. Consequently, the uncertainty
relation for this case on a lattice is \be \Delta x\cdot\Delta
k=\frac{1}{2}
\sqrt{1+\frac{t_0^{2}\cos^{2}(k_{0})t^2}{\alpha^{4}}}\;. \ee As one
can see from this expression, in general the uncertainty never
decreases as a function of time; the degree of increase depends on
the mean momentum. However, if $k_{0}=\pi/2$, the uncertainty
remains unchanged. In other words, the wave packet possesses a
solitonic behavior without showing the seemingly inevitable quantum
spreading.
We will demonstrate this fact numerically later.
This possibility is actually well known in optics,
where one seeks a medium with zero group velocity dispersion to
minimize loss.\cite{optics} Nevertheless, this appears to be less
appreciated in quantum mechanics.

This result persists in one dimension for any dispersion. That is,
one can show that some wave vector always exists for which the group
velocity dispersion is zero. In higher dimensions the situation is
not quite as simple. The result remains for nearest neighbor
hopping only. For example in three dimensions we have
$E_{k}=-2t_{0}\left[\cos(k_{x}a)+\cos(k_{y}a)+\cos(k_{z}a)\right]$,
and one readily obtains a similar result as in one dimension.
However, when next nearest neighbor hopping is included, a little
algebra shows that in general one cannot achieve conditions for zero
group velocity dispersion.

To obtain this result numerically, one diagonalizes the
tight-binding Hamiltonian, $H_{0}=-t_{0}\sum_{i}\left[
C^{+}_{i}C_{i+1}+C^{+}_{i+1}C_{i}\right]$ to obtain eigenvalues and
the corresponding eigenstates. Then one can construct a wave packet,
and evolve it in time according to the time dependent Schrodinger
equation (this procedure is described in more detail in Section IV).
The result is plotted in Fig.~1 for three different times with
$k_{0}=\pi/2$ ($k_0 = \pi/4$) in the top (bottom) panel,
respectively. A wave packet initially centered at $x_{i}=100$ moves
to $x_{i}\simeq 300$ in each panel. The uncertainty parameter
$\alpha$ is set to be $10$. The total time elapsed is different in
the two panels because of the $k_{0}$ difference (in a consistent
dimensionless time unit, $100$ for the top, and $140$ for the bottom
panel). As one can see, the wave packet with $k_{0}=\pi/2$ does not
spread and the peak height remains unchanged while the wave packet
with $k_{0}=\pi/4$ spreads and the height becomes considerably
smaller as it moves.

\begin{figure}[tp]
\begin{center}
\includegraphics[height=2.4in,width=2.4in]{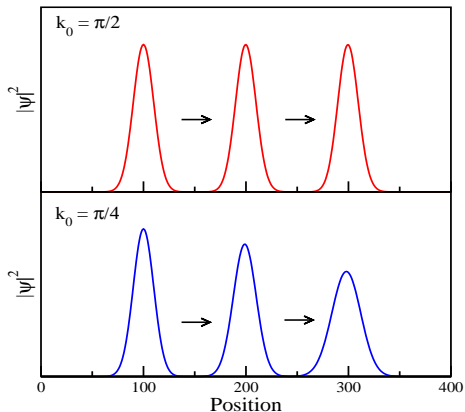}
\caption{
Time evolution of the wave packet with
$k_{0}=\pi/2$ (top) and $\pi/4$ (bottom).
The wave packet with $k_{0}=\pi/2$ does not spread
while the wave packet $k_{0}=\pi/2$ broadens as it moves.
}
\end{center}
\end{figure}

\section{quantum mechanical approach}

In the previous section we examined the free wave packet transfer in
a lattice. While construction of a packet proceeds as in the
continuum case, the tight-binding dispersion leads to a variety of
possible wave packet characteristics. In this section we examine the
consequences for scattering, in one dimension. The electronic
transfer can be explored based on the quantum mechanical approach
using the matching conditions of wave functions, or through the
transfer matrix formalism, or through direct diagonalization of the
Hamiltonian. The quantum mechanical approach and the transfer matrix
formalism are formally equivalent to each other and relatively
analytical compared with direct diagonalization, which is inherently
numerical and the subject of a later section. If there are $N$
impurities embedded in a lattice, one needs to deal with a $(2N
\times 2N)$ matrix to determine all the relevant coefficients
including $R$ and $T$ in the quantum mechanical approach. On the
other hand, the transfer matrix formalism requires manipulation of
$N$ $(2 \times 2)$ matrices. Thus, when the number of impurities is
large, say greater than $5$, the quantum mechanical approach is not
as feasible as the transfer matrix formalism.

We used the transfer matrix formalism to study symmetries of
electron-impurity scattering previously.\cite{kim06} In this paper,
we will mainly use the quantum mechanical approach in a lattice and
compare the two methods using several examples. Recall the situation
in the continuum limit: consider a Hamiltonian ${\cal
H}=-\partial^{2}_{x}/(2m) + V(x) $ where, for this illustration we
use a simple rectangular barrier potential: $V(x)=V_{0} \Theta(d/2 -
|x|)$. In order to solve the Schr{\"o}dinger equation, ${\cal
H}\psi(x) = E_{k}\psi(x)$ with $E_{k}=k^{2}/(2m)$, we write down the
most general wave function in the three regions,
\begin{center}
$
\psi(x) = \left\{
  \begin{array}{lll}
 \psi_{L}(x) = e^{ikx}+R\;e^{-ikx}& \hskip 0.25 cm \mbox{for $x \le -d/2$}
\\
 \psi_{in}(x) = A\;e^{iqx}+
               B\;e^{-iqx}& \hskip 0.25 cm \mbox{for $|x|\le d/2$}
\\
 \psi_{R}(x) = T\; e^{ikx} & \hskip 0.25 cm \mbox{for $x\ge d/2$}
  \end{array} \right.
$
\end{center}
where $q$ is determined by $q=\sqrt{2m(E_{k}-V_{0})}$. Note that
$\psi_{L,R}(x)$ describes a free electron satisfying ${\cal
H}_{0}\psi_{L,R}(x)=E\psi_{L,R}(x)$ for any value of $x$, where
${\cal H}_{0}=-\partial^{2}_{x}/(2m)$. Using the matching conditions
of $\psi(x)$ and $\partial\psi(x)/\partial x$ at $x=\pm\; d/2$, we
determine the coefficients $R$, $A$, $B$, and $T$. An
example\cite{lipkin73} of this case is the one as $d\rightarrow0$
and $V_{0}\rightarrow\infty$ while $dV_{0}$ is finite;
namely, $V(x)=dV_{0}\delta(x)$. A little algebra yields
$|T|^2=1/\left[1+\left(dV_{0}/v_{k}\right)^2\right]$ and
$|R|^2=1-|T|^2$, where the velocity $v_{k}=\partial E_{k}/\partial k
= k/m$.

In a lattice, a potential is represented by a series of
``impurities'' whose effect is to alter the on-site energy wherever
an impurity has substituted for the usual atom. This is represented
by the Hamiltonian, \be
H=-t_{0}\sum_{i}\left[C^{+}_{i}C_{i+1}+C^{+}_{i+1}C_{i}\right]+
\sum_{i\in {\cal I}}U_{i}C^{+}_{i}C_{i}\;, \label{ham_second} \ee
where $t_{0}$ is the hopping amplitude as before, $C^{+}_{i}$
creates an electron at a site $i$, and $U_{i}$ is a scalar potential
at site $i$; the set of "impurities" spans a number ${\cal I} =
\{0,1,2,\cdots,I\}$. Note that $U_{i}$ can be positive (repulsive),
negative (attractive), or zero. In this way one can construct any
shape potential one wishes (on a scale much greater than the lattice
spacing). Since we consider only scalar potentials, and we are
treating the one electron problem here, we ignore the spin index of
the electrons. The wave function defined only on the lattice sites
can now be written as a piecewise function over $I + 3$ regions
since the number of impurities is $I+1$. That is, $|\psi\ra =
\sum_{j}\psi(x_{j})C^{+}_{j}|0\ra$, where
\begin{center}
$
\psi(x_{j}) = \left\{
  \begin{array}{lll}
 \psi_{L}(x_{j}) = e^{ikx_{j}}+R\;e^{-ikx_{j}}& \hskip 0.25 cm
\mbox{for $j < 0$}
\\
 \psi_{j}(x_{j}) = A_{j}\;e^{iq_{j}x_{j}}+
               B_{j}\;e^{-iq_{j}x_{j}}& \hskip 0.25 cm
\mbox{for $j\in{\cal I}$}
\\
 \psi_{R}(x_{j}) = T\; e^{ikx_{j}} & \hskip 0.25 cm
\mbox{for $j > I$}
\end{array} \right.
$
\end{center}
and $|0\ra$ represents the vacuum, namely, the state with an empty
lattice. Since we set the lattice constant to be unity, the
displacement $x_{i} =a\cdot i=i$. The coefficients $R$, $A_{i}$,
$B_{i}$, and $T$ are to be determined by matching conditions at
$i\in{\cal I}$. Note that $q_{i}$ will be obtained within the same
calculation even though we can already guess that
$E_{k}=E_{q_i}+U_{i}$. What are the matching conditions ? Like the
continuum limit one first demands continuity of the wave function at
each site. Therefore, $\la0|C_{j+}|\psi\ra=\la0|C_{j-}|\psi\ra$,
where $j+$ ($j-$) means just to the right (left) of site $j$.
However, the second condition in the continuum limit requires
continuity of the {\em derivative} of the wave functions at each
site. One can see this directly from the Schr{\"o}dinger equation
through an integration of the second order differential equation.
However, the second quantized form of the Hamiltonian written in
Eq.~(\ref{ham_second}) contains no derivatives, so clearly this
procedure is not an option. The correct procedure is as
follows.\cite{hirsch94} One first writes down the Schr{\"o}dinger
equation projected onto each site,
$\la0|C_{j}H|\psi\ra=\la0|C_{j}E|\psi\ra$ for $j=0,1,\cdots,I$. Then
the two conditions, expressed for each site, can be written \bea
\psi(j+0^{+})&=&\psi(j+0^{-}) \label{match}
\\
-t_{0}\left[\psi(j+1)+\psi(j-1)\right]&+&U_{j}\psi(j)=E\psi(j)\;.
\label{second_eq}
\eea
As mentioned before, Eq.~(\ref{match})
implies the wave function is continuous at each site.
Eq.~(\ref{second_eq}) is the Schr{\"o}dinger equation at each site;
however, close inspection shows that on the left-hand side the wave
function is required from two different "pieces" in the domain
(assuming that $U_j$ is non-zero. But we would like the
Schr{\"o}dinger equation for non-interacting electrons to be
satisfied, {\em with the same eigenvalue},  by the wave function on
any given "piece" even when extended beyond the domain of validity
of that wave function. For example, for $j = 0$ we demand that
$\psi_L$ satisfy the equation,
$-t_{0}\left[\psi_{L}(+1)+\psi_{L}(-1)\right]=E\psi_{L}(0)$.
Note that we have used the wave function $\psi_L$ at location $+1$,
even though it was originally defined
only for sites 0 or below. Moreover, we require that this equation
be satisfied with the same eigenvalue, $E$. Hence, by judicious
adding and subtracting of a wave function to Eq.~(\ref{second_eq})
at each impurity site, we arrive at, for $j=0$ and $j = I$:
\bwt
\bea
-t_{0}\psi_{1}(1)+t_{0}\psi_{L}(1)+U_{0}\psi_{L}(0)&=&0\hskip
0.25cm \mbox{for $j=0$}
\\
-t_{0}\psi_{I}(I-1)+t_{0}\psi_{R}(I-1)+U_{I}\psi_{R}(I)&=&0\hskip
0.25cm \mbox{for $j=I$}\;.
\eea
\ewt
Similar equations apply for the impurity sites in between. These
can now be solved for the unknown coefficients along with the "matching
equations" [Eq.~(\ref{match})].

Let us consider a two impurity case as an example. Assume two
impurities with $U_{0}$ and $U_{1}$ are embedded at $j=0$ and $1$,
respectively, in a lattice; one needs to introduce a wave function
as follows:
\begin{center}
$
\psi(x_{j}) = \left\{
  \begin{array}{llll}
 \psi_{L}(x_{j}) = e^{ikx_{j}}+R\;e^{-ikx_{j}}
\\
 \psi_{0}(x_{j}) = A_{0}\;e^{iq_{0}x_{j}}+
               B_{0}\;e^{-iq_{0}x_{j}}
\\
 \psi_{1}(x_{j}) = A_{1}\;e^{iq_{1}x_{j}}+
               B_{1}\;e^{-iq_{1}x_{j}}
\\
 \psi_{R}(x_{j}) = T\; e^{ikx_{j}}.
  \end{array} \right .
$
\end{center}
The coefficients such as $R$ and $T$, and the momenta $q_{0}$, and $q_{1}$
can be determined
by solving the continuity equations
$\psi_{L}(-1)=\psi_{1}(-1),\;
\psi_{L}(0)=\psi_{1}(0),\;
\psi_{1}(0)=\psi_{2}(0),\;
\psi_{1}(1)=\psi_{2}(1),\;
\psi_{2}(1)=\psi_{R}(1),\;
\psi_{1}(2)=\psi_{R}(2)$,
and the Schr{\"o}dinger equations
\bea
-\psi_{1}(1)+\psi_{L}(1)&+&U_{0}\psi_{L}(0)=0
\nonumber\\
-\psi_{1}(0)+\psi_{R}(0)&+&U_{1}\psi_{R}(1)=0\;,
\eea
where we set the
nearest neighbor hopping amplitude $t_{0}$ to be unity for
simplicity. It is straightforward to show that
\bea
T&=&\frac{2i\sin(k)}
{2i\sin(k)-(U_{0}+U_{1})-U_{0}U_{1}e^{ik}}
\label{T_2}
\\
R&=&\frac{U_{0}+U_{1}e^{2ik}+U_{0}U_{1}e^{ik}}
{2i\sin(k)-(U_{0}+U_{1})-U_{0}U_{1}e^{ik}}.
\label{R_2} \eea%
Using the Schr{\"o}dinger equations, one can show that
$E_{k}=E_{q_i}+U_{i}$ $(i=1,2)$. This result will be used to compare
the quantum mechanical approach and the transfer matrix formalism. A
case of two impurities with same potentials $(U_{0}=U_{1})$ has been
well studied in a context of the random dimer
model.\cite{dunlap,wu1,wu}%

In Fig.~2, we plot $|T|^2$ for two impurities side by side using
Eq.~(\ref{T_2}). Fig.~2(a) to Fig.~2(c) are contour plots of
$|T(E,U_{1})|^{2}$ for given values of $U_{0}=-1,\;1$, and $1.8$,
respectively. Note that Fig.~2(a) and Fig.~2(b) illustrate a
symmetry, $|T(E,\;U_{0},\;U_{1})|^{2}=
|T(-E,\;-U_{0},\;-U_{1})|^{2}$, which we showed in a previous
paper.\cite{kim06} When we increase $U_{0}$ further, $|T|^{2}$
becomes suppressed considerably as shown in Fig.~2(c). We also plot
$|T|^{2}$ vs. $E$ in Fig.~2(d) for various values of $U_{1}$ while
$U_{0}$ is fixed to be $1.0$. In fact, Fig.~2(d) can be read off
from Fig.~2(b) along a line of the corresponding value of $U_{1}$.
In the case of two impurities, a transmission resonance
$(|T|^{2}=1)$ occurs when $E=U_{0}=U_{1}$ and $|E|\le2$. As one can
see in Fig.~2(d), a transmission resonance $(|T|^{2}=1)$ occurs at
$E=1$ for $U_{0}=U_{1}=1$.%
\vskip 1.5cm
\bwt
\begin{figure}[tp]
\begin{center}
\includegraphics[height=6.5 in,width=5 in,angle=-90]{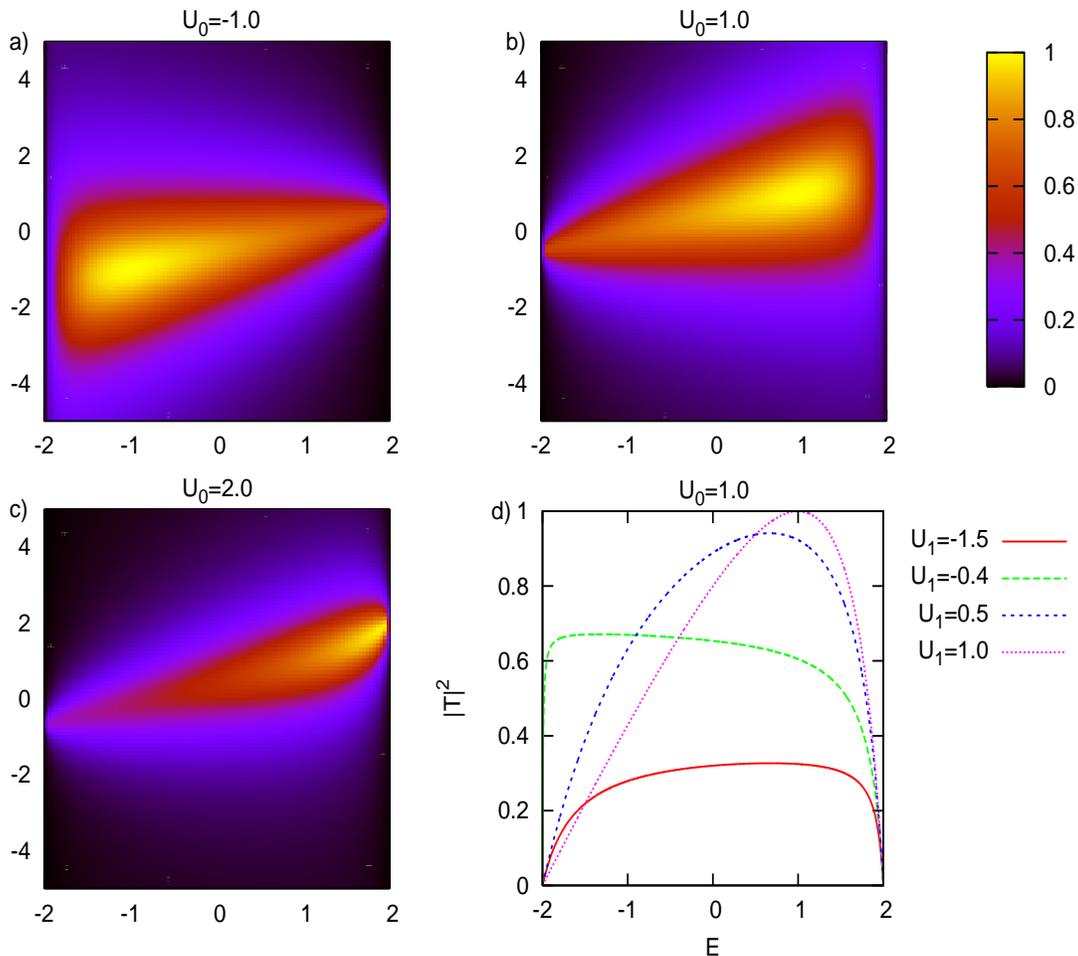}
\caption{ 
$|T|^2$ for two impurities using Eq.~(\ref{T_2}). Fig.~2(a) to Fig.~2(c) are contour plots of $T(E,U_{1})|^{2}$ for
given values of $U_{0}=-1,\;1$, and $1.8$, respectively. Fig.~2(d) is $|T|^{2}$ as a function of $E$ for various values of $U_{1}$ with
$U_{0}=1.0$. }
\end{center}
\end{figure}

\begin{figure}[tp]
\begin{center}
\includegraphics[height=6.5 in,width=5 in,angle=-90]{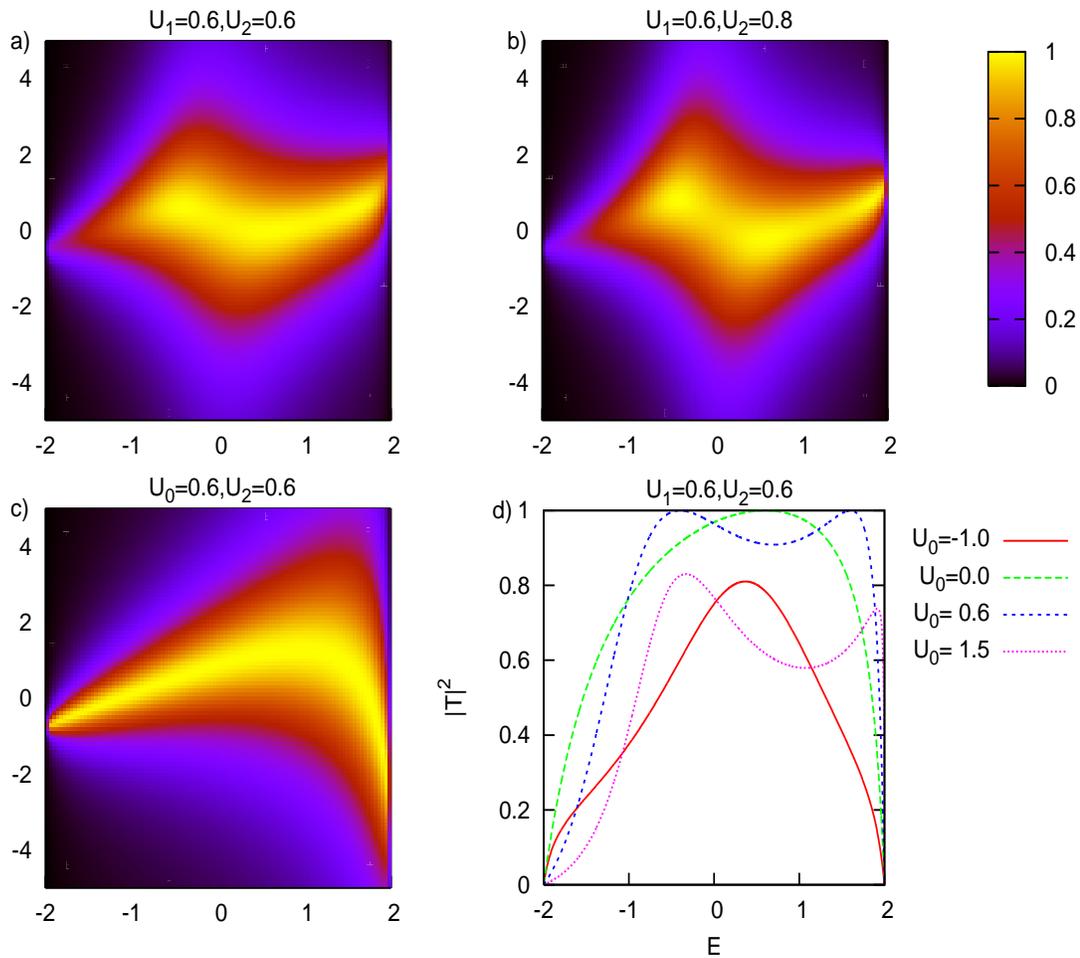}
\caption{
$|T|^{2}$ for three impurities using Eq.~(\ref{T_3}).
Fig.~3(a) to Fig.~3(c) are contour plots while Fig.~3(d) is for $|T|^{2}$
vs. $E$. Fig.~3(a) is for $|T(E,U_{0})|^{2}$ with $U_{1}=U_{2}=0.6$.
Fig.~3(b) is also for $|T(E,U_{0})|^{2}$ but with
$U_{1}=0.6$ and $U_{2}=0.8$.
Fig.~3(c) shows $|T(E,U_{1})|^{2}$ with $U_{0}=U_{2}=0.6$.
Fig.~3(d) is for $|T|^{2}$ vs. $E$, which can be read off from Fig.~3(a)
along a line of the corresponding value of $U_{0}$.
}
\end{center}
\end{figure}
\ewt

For three impurities with $U_{0}$, $U_{1}$, and $U_{2}$ at $j=0$,
$1$ and $2$, respectively, a similar calculation gives the
transmission amplitude
\bwt%
\be%
T=\frac{-2i\sin(k)}
{(U_{1}-E_{k})+e^{ik}\left[(U_{0}+U_{2})(U_{1}-E_{k})-2\right]
+e^{2ik}\left[U_{0}U_{2}(U_{1}-E_{k})-(U_{0}+U_{2})\right] }.%
\label{T_3}%
\ee%
\ewt%
It is easy to show that Eq.~(\ref{T_3}) reduces to Eq.~(\ref{T_2})
as $U_{0}$ or $U_{2}$ goes to zero. Giri {\it et al.}\cite{giri}
studied this case in detail and classified the parameter values into
five cases, for which perfect transmission can be obtained. The
first is the dimer case, and the second only applies if the hopping
parameters in the impurity region are different, so here we will
consider only the remaining three cases. Let us briefly describe the
three cases using our notation. In case (III), $U_{0}=U_{1}=U_{2}$
and $E=U_{0}\pm 1$. This is the straightforward extension of the
symmetric dimer case to a symmetric trimer, but note that now the
potential for two solutions exists. In case (IV), if
$U_{0}+U_{2}=U_{1}$ then resonance is obtained at an energy,
$E=U_{1}$, regardless of the values of $U_0$ and $U_2$. We will note
an example below. Case (V) is the most interesting; in this instance
$U_{0}=U_{2}$ and $U_1$ is arbitrary. Then resonance is obtained for
energies,
\be%
E=U_{1}+\frac{1}{2}\left[
U_{1}-U_{0}\pm\sqrt{(U_{1}-U_{0})^{2}-4(U_{1}/U_{0}-2)}\right],
\label{case5}%
\ee%
provided that $(U_{1}-U_{0})^{2}\ge 4(U_{1}/U_{0}-2)$ and of course
$-2\le E \le 2$. We plot $|T|^{2}$ for three impurities sitting side
by side in Fig.~3. Fig.~3(a) to Fig.~3(c) are contour plots while
Fig.~3(d) is for $|T|^{2}$ vs. $E$. Fig.~3(a) is for
$|T(E,U_{0})|^{2}$ with $U_{1}=U_{2}=0.6$ and illustrates both case
(III) and case (IV). For example, if $U_{0}=0.6$, $|T|^{2}=1$ at
$E=1.6$ and $-0.4$. This belongs to case (III). On the other hand,
since $U_{1}=U_{2}$, then when $U_{0}=0$, $|T|^{2}=1$ at $E=0.6$,
which belongs to case (IV). In Fig.~3(b), which is also a contour
plot of $|T(E,U_{0})|^{2}$ but with $U_{1}=0.6$ and $U_{2}=0.8$, we
still have case (IV) when $U_{0}=-0.2$ and $E=0.6$. On the other
hand, when $U_{0}=0.8$, $|T|^{2}=1$ at $E=1.82$ and $-0.42$, which
belongs to case (V). Fig.~3(c) is primarily an example of case (V)
since $U_{0}=U_{2}$. The condition $(U_{1}-U_{0})^{2}\ge
4(U_{1}/U_{0}-2)$ requires $U_{1}<1.27$ for $U_{0}=0.6$. Since
$|E|\le2$, one finds that there exist two solutions for $|T|^{2}=1$
if $-0.75 \le U_{1}\le1.27$; however, for $-3<U_{1}<-0.75$ there is
only one solution as can be observed in Fig.~3(c). Also note that
Fig.~3(c) contains as a special case the `spread-out' dimer, where
the middle of the trimer has no impurity ($U_1 = 0$), so the
remaining dimer is now separated by two lattice spacings. This is a
special example of Giri et al.'s case (V), and is also a special
case of the arbitrarily spread-out dimer considered by us previously
\cite{kim06}.

Fig.~3(d) is for $|T|^{2}$ vs. $E$, which can be read off from
Fig.~3(a) along a line of the corresponding value of $U_{0}$. Note
the two solutions for perfect transmission for the symmetric trimer
case (dashed blue curve, $U_0 = 0.6$). Also shown is a dimer case
(dashed green curve), when $U_0 = 0$, for which $E = 0.6$ yields
resonant transmission. The standard one dimensional plots are quite
clear; on the other hand the contour plots provide a feel for how
strongly the transmission remains as parameters vary away from the
resonant condition.

\section{transfer matrix formalism}

In the transfer matrix formalism,\cite{gilmore} we write the
Schr{\"o}dinger equation Eq.~(\ref{second_eq}) in the matrix form as
follows: \be \left(\begin{array}{c}
\psi_{j+1}\\
\psi_{j}
\end{array}\right)
= M_{j}
\left(\begin{array}{c} \psi_{j}\\
\psi_{j-1}
\end{array}\right)
\ee where $M_{j}=\left(\begin{array}{cc}
U_{j}-E_{k} & -1\\
1 & 0
\end{array}\right)$,
which is a unimodular matrix and associated with
an impurity at the site $j$.
The wave functions $\psi_{L}$ (for $i<1$) and $\psi_{R}$ (for $i>N$)
are $\psi_{L}=e^{ikx_{i}}+R\;e^{-ikx_{i}}$ and
$\psi_{R}=T\;e^{ikx_{i}}$. Using the
transfer matrix formalism, one can express the coefficients $R$ and
$T$ in terms of $k$, $U_{i}$, and $E$ as follows:
\be
\left(\begin{array}{c}
{\cal T}\\
i{\cal T}
\end{array}\right)
=P
\left(\begin{array}{c}
1+{\cal R}\\
i(1-{\cal R})
\end{array}\right)\;,
\label{tr_eq}
\ee
where $P=S^{-1}MS$ with $S=\left(\begin{array}{cc}
\cos(k) & \sin(k)\\
1 & 0
\end{array}\right)$, and $M=M_{N}M_{N-1}\cdots M_{1}$.
Solving
Eq.~(\ref{tr_eq}), one can obtain\cite{burrow}
\bea
{\cal T}&=&\frac{2i}{i\left(P_{11}+P_{22}\right)+P_{12}-P_{21}}
\\
{\cal R}&=&\frac{P_{12}+P_{21}-i\left(P_{11}-P_{22}\right)}
{i\left(P_{11}+P_{22}\right)+P_{12}-P_{21}}\;. \eea It is
instructive to compare the quantum mechanical approach and the
transfer formalism using an example. Let us consider two impurities
residing side by side. In this instance we know that the
transmission and reflection amplitude are Eqs.~(\ref{T_2}) and
(\ref{R_2}). In the transfer matrix formalism, one needs to
calculate $P=S^{-1}M_{1}M_{0}S$ to obtain the transmission amplitude
\bwt \be {\cal T}=\frac{2i\sin(k)}
{i\sin(k)\left[(U_{0}-E_{k})(U_{1}-E_{k})-2\right]+U_{0}+U_{1}
-2E_{k}-\cos(k)(U_{0}-E_{k})(U_{1}-E_{k})} \label{T_2_tr} \ee
 \ewt
Note that $T$ [Eq.~(\ref{T_2})] is not identical with ${\cal T}$
[Eq.~(\ref{T_2_tr})]. However, while not obvious, they merely differ
by a phase factor and their magnitudes are the same: ${\cal T}=
e^{2ik}T$. Clearly the transfer matrix method `keeps track' of the
two lattice spacings traversed as the particle is transmitted to the
other side. A similar relation holds for the reflection amplitude.
Consequently, the transmission (reflection) probabilities are
identical; $|T|^{2} = |{\cal T}|^{2}$ and $|R|^{2} = |{\cal R}|^{2}$
in the two formalisms.

In previous work\cite{kim06} using the transfer matrix formalism, we
derived a relation between ${\cal R}$ and ${\cal R}'$, where ${\cal
R}'$ is the reflection amplitude for the reverse impurity
configuration. It is \be \frac{{\cal T}^{*}\left({\cal R}'-{\cal
R}\right)} {{\cal T}\left({\cal R}'-{\cal R}\right)^{*}}=e^{2ik}\;.
\ee In the example of the two impurities, ${\cal R}'$ is given by
${\cal R}$ with $U_{0}$ and $U_{1}$ exchanged. Using
Eqs.~(\ref{T_2}) and (\ref{R_2}), one can show that $T$ and $R$ also
satisfy the same relation; namely, \be
\frac{T^{*}\left(R'-R\right)}{T\left(R'-R\right)^{*}}=e^{2ik} \ee
Introducing a phase difference between $R$ and $R'$ such as
$R'=e^{\delta}R$, we obtain \be
e^{i\delta}=-e^{2ik}\;\frac{R^{*}T}{RT^{*}}\;. \ee For the case of
two impurities we considered, the phase difference $\delta$ can be
determined by \be \tan(\delta-2k)=\frac{2\mu\nu}{\nu^{2}-\mu^{2}}
\label{delta_2} \ee where \bea
\mu&=&U_{0}+U_{1}\cos(2k)+U_{0}U_{1}\cos(k)
\\
\nu&=&U_{1}\sin(2k)+U_{0}U_{1}\sin(k)
\eea
Later, we will show that the derivative of $\delta$ with respect to $k$
can be attributed to a phase shift (or time delay) of wave functions
for the reverse
configuration of impurities.

\section{numerical diagonalization and phase shift of wave packets}

We have already alluded to the numerical approach for a free wave
packet transfer in section II. Since the operator $C^{+}_{i}$
creates an electron at site $i$, the initial wave packet can be
written as $|\Psi(0)\ra = \sum_{i}\varphi(x_{i},0)C^{+}_{i}|0\ra$,
where \be \varphi(x_{i},0)=\frac{1}{(2\pi
a^{2})^{1/4}}e^{ik_{0}(x_{i}-x_{0})}
e^{-\frac{1}{4}(x_{i}-x_{0})^{2}/\alpha^{2}}. \label{initial_second}
\ee As mentioned in Sec.~II, $x_{0}$ is the mean position, $k_{0}$
is the mean momentum, and $\alpha$ is the initial uncertainty
associated with the position. If $\alpha$ is much larger than the
size of the potential region, say $I$, the wave packet acts like a
plane wave when it is scattered off the potential. To see the real
time propagation of a wave packet in a lattice with $N_{0}$ sites in
total, one needs to diagonalize the Hamiltonian
Eq.~(\ref{ham_second}), which is an $(N_{0}\times N_{0})$ matrix.
Since the impurity potentials are real, the Hamiltonian matrix is
real and symmetric. The numerical diagonalization is done using the
expert driver DSYEVX contained in the LAPACK package, which provides
either selected eigenvalues and eigenvectors or the entire spectrum.
Using the eigenstates $|n\ra$ and eigenvalues $\epsilon_{n}$
obtained from the diagonalization, one can express the wave packet
at time $t$ as follows: \be |\Psi(t)\ra = \sum^{N_{0}}_{n=1}|n\ra\la
n|\Psi(0)\ra e^{-i\epsilon_{n}t}\;. \ee The wave packet initially at
$x_{0}$ moves to the potentials and scatters off  impurities. In
general, the wave packet is partially reflected and partially
transmitted. The mathematical definitions of the reflection and
transmission probabilities are
$|R|^{2}=\sum_{i<0}|\varphi(x_{i},t)|^{2}$ and
$|T|^{2}=\sum_{i>I}|\varphi(x_{i},t)|^{2}$, respectively as
$t\rightarrow\infty$.

Let us consider the two impurity case again to illustrate the time
evolution of a wave packet in the presence of impurities in a
lattice. The impurity potentials are set to be $U_{0} = 1$ and
$U_{1} = 3$ in units of the hopping constant $t_{0}$. We consider
wave packets with average momentum that varies from $k_{0}=0.3\pi$
to $0.9\pi$. The time elapsed for the scattering processes to
`finish' depends on $k_{0}$; for example, for $k_{0}=0.6\pi$ it is
$160$ in our dimensionless time unit, while for $k_{0}=0.8\pi$, it
would be $240$. We consider two impurity configurations, [I] and
[II]. For [I], we have $(U_{0},U_{1})$ and for [II],
$(U_{1},U_{0})$. As we discussed earlier, the reflection amplitude
will differ correspondingly; namely, $R$ for [I] while $R'$ for
[II]. Since $T=T'$, there is no phase shift for the transmitted wave
packet as shown in Ref. \onlinecite{kim06}. On the other hand, the
phase difference $\delta$ induces the phase shift for the reflected
wave packets. This can be explained as follows: Consider a wave
packet $\psi(x,0)$ moving with $k_{0}$ to an impurity region \be
\psi(x,0)=\int{}dk\; g(k)\;e^{ik_{0}(x-x_{0})} \ee where $g(k)\sim
e^{-\alpha(k-k_{0})^{2}}$. After the wave packet scatters completely
off the impurities at time $t_{s}$, the wave function at $t$ after
$t_{s}$ would be \bea \psi_{[I]}(x,t)&=&\int{}dk\;
g_{R}(k)e^{-ik_{0}(x-x_{R})}e^{-iE_{k}t}
\nonumber\\
&+&\int{}dk\; g_{T}(k)e^{ik_{0}(x-x_{T})}e^{-iE_{k}t}
\eea
where an elastic scattering is assumed,
$g_{R}\sim R(k)g(k)$ while $g_{T}\sim T(k)g(k)$, and
$x_{R} (x_{T})$ is the mean position of the reflected (transmitted)
wave packet at $t_{s}$.
For the reverse configuration [II],
\bea
\psi_{[II]}(x,t)&=&\int{}dk\; g_{R'}(k)e^{-ik_{0}(x-x_{R})}e^{-iE_{k}t}
\nonumber\\
&+&\int{}dk\; g_{T'}(k)e^{ik_{0}(x-x_{T})}e^{-iE_{k}t} \eea Since
$R'=R\;e^{i\delta}$, the reflected part of $\psi_{[II]}(x,t)$
becomes \be \psi_{[II],R}(x,t)\sim \int{}dk\;
R(k)e^{-\alpha(k-k_{0})^{2}}e^{-ik_{0}(x-x_{R})-iE_{k}t+i\delta} \ee
In order to see the phase shift of the reflected wave packets, one
needs to calculate $|\psi_{[II],R}|^{2}$. Expanding $E(k)$ and
$\delta(k)$ around $k_{0}$ and some algebra yields \be
|\psi_{[II],R}|^{2}\sim exp\left[-\frac
{\left(x-x_{R}+v_{0}t-\partial_{k}\delta_{0}\right)^{2}}
{2\left\{\alpha^{2}+t^{2}(\partial^{2}_{k}E_{0}-\partial^{2}_{k}
\delta_{0})^{2}/4\right\}} \right] \label{phase_shift} \ee where
$v_{0}=\partial_{k}E_{0}$. Rigorously speaking, this approximation
is valid when $R(k)\simeq R(k_{0})$. Note that the term with $E_{0}$
and $\delta_{0}$ disappears when $|\psi_{[II],R}|^{2}$ is calculated
and the position of the scattered wave is $x_{R}-v_{0}t$.
Eq.~(\ref{phase_shift}) indicates that the phase shift of the
reflected wave packets is determined by $\partial_{k}\delta_{0}$,
and the spreading depends not only on $\partial^{2}_{k}E_{0}$ but
also on $\partial^{2}_{k}\delta_{0}$. To be specific we define a
phase shift as the difference between the two reflected wave packets
for [I] and [II] at their half width as in Ref. \onlinecite{kim06}.

\begin{figure}[tp]
\begin{center}
\includegraphics[height=2.4in,width=2.4in]{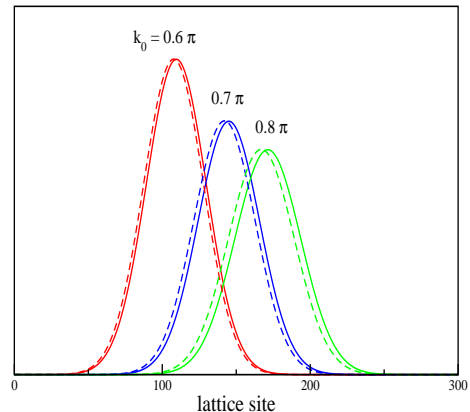}
\caption{ Phase shift of the reflected wave packets in a lattice
with two impurities embedded side by side with $U_{0} = 1$ and
$U_{1} = 3$ in units of the hopping constant. Initially, the wave
packet is at $x_{0}=150$. The two impurities are at $300$ and $301$.
The uncertainty parameter $\alpha$ is set to be $20$. For
$|\psi_{[I],R}|^{2}$ (solid curves), the impurity configuration is
$(U_{0},U_{1})$ while for $|\psi_{[II],R}|^{2}$ (dashed curves), it
is $(U_{1},U_{0})$. }
\end{center}
\end{figure}

In Fig.~4, we plot the reflected wave packets for [I] (solid curves)
and [II] (dashed curves) in the case of two impurities with $U_{0} =
1$ and $U_{1} = 3$. The average momenta for the three wave packets
shown are $k_{0} = 0.6\pi$, $0.7\pi$, and $0.8\pi$. As one can see
clearly, the phase shift between $|\psi_{[I],R}|^{2}$ (solid curves)
and $|\psi_{[II],R}|^{2}$ (dashed curves) depends on $k_{0}$. The
phase shift for $k_{0}=0.6\pi$ is not significant while it becomes
bigger as $k_{0}$ increases to $0.8\pi$. We also plot the phase
difference, its derivative, and the phase shift of the reflected
wave packets in Fig.~5. One may use Eq.~(\ref{delta_2}) to obtain
the phase difference and its derivative in this simple case.
However, when we consider many impurities we need to use the
transfer matrix formalism. As shown in the plot, the derivative of
$\delta$ and the phase shift obtained numerically are in excellent
agreement.

\begin{figure}[tp]
\begin{center}
\includegraphics[height=2.4in,width=2.4in]{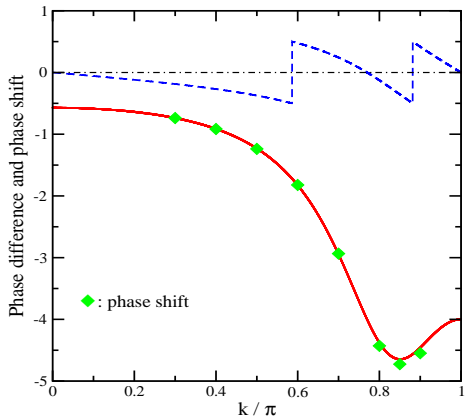}
\caption{ Phase difference (blue dashed curve), its derivative (red
solid curve), and the phase shift obtained from the wave packet
simulations (green diamond symbols). The phase shift is measure by
the difference between the reflected wave packets $\psi_{[I],R}|$
and $\psi_{[II],R}|$ at their half width. The impurity potentials
are $U_{0}=1$ and $U_{1}=3$. The phase shift and
$\partial\delta/\partial k$ agree with one another very well. }
\end{center}
\end{figure}

\section{spin flip scattering}

So far we have discussed scalar potential scattering on a lattice.
In this section we describe how to study spin flip scattering on a
lattice. Let us consider the Hamiltonian \be
H=-t_{0}\sum_{<i,j>\sigma}C^{+}_{i\sigma}C_{j\sigma}
-2J_{0}\sum_{l}{\bf\sigma}_{l}\cdot{\bf S}_{l} -2J_{1}\sum_{l}{\bf
S}_{l}\cdot{\bf S}_{l+1} \ee where $C^{+}_{i\sigma}$ creates an
electron with a spin $\sigma$ at a site $i$, ${\bf S}_{l}$ is a spin
operator located at a site $l$, $t_{0}$ is a hopping amplitude
between the nearest neighbor sites, and $J_{0}$ is the coupling
between an electron and a local spin. The electron-spin coupling is
assumed here to be a purely local (i.e. on-site) interaction.
$J_{1}$ is a (Heisenberg exchange) coupling between two neighbouring
spins. This model can be used to understand the spin transfer
dynamics between an itinerant electron and a ferromagnetic spin
chain with $N_{s}$ local spins $(S=1/2)$ arranged in a one
dimensional lattice. To study the spin chain, one can extend the
quantum mechanical approach as in Ref. \onlinecite{kim05}, or one
can extend the transfer matrix formalism.\cite{avishai} Here we
follow the diagonalization method to see the time evolution of the
spins.

Suppose we send a wave packet representing an electron with spin
aligned in the $+Z$ direction towards a ferromagnetic spin chain
where all spins are aligned in the $-Z$ direction. Such a state of
the chain will be denoted by $\Gs$. The incoming wave packet can be
constructed as follows:%
\be%
|\psi(0)\ra = \sum_{i}\varphi(x_{i},0)C^{+}_{i\up}|0\ra%
\ee%
Then, the initial wave function of the total system including the
incoming electron and the chain is
$|\Psi(0)\rangle=|\psi(0)\rangle\Gs$. We introduce the total spin
operator ${\bf J} = {\bf\sigma}+\sum_{l}{\bf S}_{l}$. The $Z$
component, $J_{z}$, of the total spin is conserved. Hereafter we
assume spin $1/2$ for both the electron and the spins, for
simplicity. Since the initial value of $J_{z}$ is $(1-N_{s})/2$, the
possible spin bases would be $\ups\Gs$ and $\dns S_{l+}\Gs$, where
$S_{l+}$ flips the local spin at $l$ in the chain. Alternative spin
bases could be used by utilizing the total spin and its Z component.
Including the location of the incoming electron, the bases states we
use are $C^{+}_{i\up}\Vs\Gs$ and $C^{+}_{i\dn}\Vs S_{l+}\Gs$. Now,
the Hamiltonian becomes an $N_{0}(N_{s}+1)\times N_{0}(N_{s}+1)$
matrix. Note that the dimension of the Hamiltonian depends only on
$N$ and $N_{s}$ and does not depend on the locations of the local
spins. Even if we include impurities, the dimension of the
Hamiltonian remains unchanged. To construct the Hamiltonian matrix,
we need to calculate each component of the matrix. For example, \bea
\Gc\Vc C_{j\up}\left(-2J_{0}{\bf \sigma}_{l}\cdot{\bf S}_{l}\right)
C^{+}_{i\up}\Vs\Gs = \frac{J_{0}}{2}\delta_{j,l}\delta_{i,l}
\nonumber\\
\Gc\Vc C_{j\up}\left(-2J_{1}{\bf S}_{l}\cdot{\bf S}_{l+1}\right)
C^{+}_{i\up}\Vs\Gs = -\frac{J_{1}}{2}\delta_{i,j}\;.
\nonumber
\eea

We therefore solve the eigenvalue problem:
$H|\eta_{j}\rangle=E_{j}|\eta_{j}\rangle$, where $j=1$, $2,\cdots
N_{0}(N_{s}+1)$. Then, we represent the time dependent total state
using eigenvalues and eigenstates as follows:
$|\psi(t)\rangle=\sum_{j}|\eta_{j}\rangle
\langle\eta_{j}|\psi(0)\rangle e^{-iE_{j}t}$. Alternatively, the
total state at $t$ can be expressed as \be
|\psi(t)\rangle=\sum^{N_{0}}_{i=1}\psi_{i}(t)C^{+}_{i\up}\Vs\Gs
+\sum^{N_{s}}_{l=1}\sum^{N_{0}}_{i=1}\psi_{l,i}(t)C^{+}_{i\dn}\Vs
S_{l+}\Gs\;. \ee Using this expression we can investigate the
dynamics of a particular spin or the sum of all spins in the chain.
For example, \be \langle\psi(t)|S_{lz}|\psi(t)\rangle=
-\frac{1}{2}\sum_{i'}|\psi_{i'}|^{2}+
\frac{1}{2}\sum_{l',i'}|\psi_{l',i'}|^{2}\left(2\delta_{l,l'}-1\right)\;,
\ee and the total local spin is $\sum_{l}\langle {\bf
S}_{l}(t)\rangle$. Since at $t=0$, $\psi_{l,i}=0$ and
$\sum_{i}|\psi_{i}|^{2}=1$, $\langle S_{lz}(0)\rangle = -1/2$ is
assumed. Thus, $\langle S_{z}(0)\rangle = \sum_{l}\langle
S_{lz}(0)\rangle = -N_{s}/2$ and $\langle J_{z}(0)\rangle =
(1-N_{s})/2$.

By way of an example, consider a chain of three spins,
ferromagnetically coupled with strength $J_1$. As an electron
impinges on the three spin system, they interact with the electronic
spin and change their states. In Fig.~6, we first plot the
expectation value of $S_{z}$ for ${\bf S}_{i}$ $(i=1,2,3)$, and
${\bf S}_{tot}=\sum^{3}_{i=1}{\bf S}_{i}$ for the uncoupled spin
case, i.e. with $J_1 = 0$. The electron is coupled to each spin with
strength $J_{0}=2$, in units of $t_{0}$. Since there is no coupling
between two nearest spins in this instance, each spin evolves
independently as a function of time. The spin transfer from the
incoming electron to the local spins occurs mostly for ${\bf S}_{1}$
while it is minimal for ${\bf S}_{3}$.

On the other hand, when the spins are Heisenberg coupled with
strength $J_{1} = 1$ all three spins participate in the spin-flip
scattering with the incoming electron almost to the same degree, as
shown in Fig.~7. Interestingly, even though there is an obvious
asymmetry (the electron strikes the first spin first) the time
evolution of the first $(S_{1})$ and the third $(S_{3})$ spins are
almost identical.

\begin{figure}[tp]
\begin{center}
\includegraphics[height=2.4in,width=2.4in]{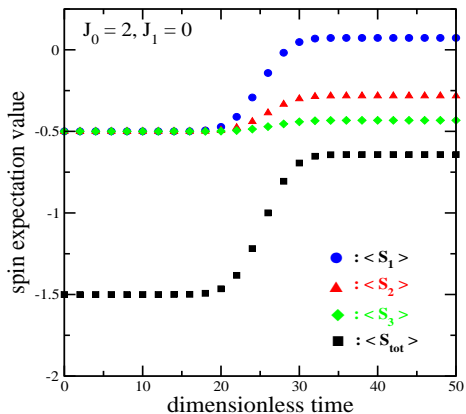}
\caption{The spin expectation value $\langle S_{z} \rangle$
for three spins $S_{i}$ $(i=1,2,3)$ and the total spin $S_{tot}$.
The coupling parameters are
$J_{0} = 2$ and $J_{1} = 0$.
}
\end{center}
\end{figure}

\begin{figure}[tp]
\begin{center}
\includegraphics[height=2.4in,width=2.4in]{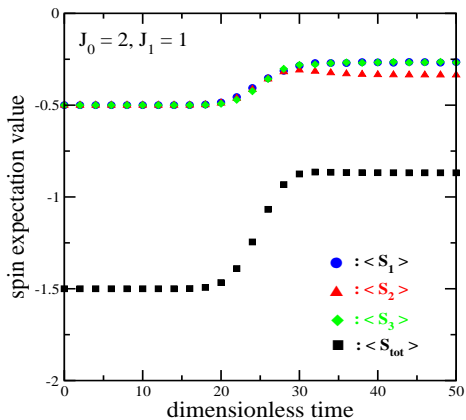}
\caption{The spin expectation value $\langle S_{z} \rangle$
for three spins $S_{i}$ $(i=1,2,3)$ and the total spin $S_{tot}$.
The coupling parameters are
$J_{0} = 2$ and $J_{1} = 1$.
}
\end{center}
\end{figure}

\section{conclusions}

In this paper we have described three rather general methods to
solve scattering problems on a lattice. The first two are based on
single electron-impurity scattering problems, and tell us about the
end result of the scattering process. We examined several examples
of impurity scattering. In particular, following Giri et al.
\cite{giri}, we explored parameter regions where near resonant
transmission is expected to occur for the random trimer model. As is
evident from Fig.~3, there are regions in parameter space where near
resonant transmission will remain as imperfections arise in the
trimer itself. It would be very interesting to examine the width of
the resonance for N such trimers arranged randomly on an infinite
lattice, as N increases.

The third method developed in this paper is numerically intensive,
but allows for solution of completely general problems, including
those involving spin-flip scattering, as explored briefly in the
last section. Moreover, one can monitor the time-dependence of the
wave function amplitudes throughout the scattering event. This will
be of interest as experimental methods evolve to allow more
controlled temporally and spatially resolved measurements. We
presented, by way of example the case of a Heisenberg chain of
spins, interacting with an itinerant electron (spin current). A
natural extension of this calculation could be to include more
spins, so as to model an actual magnetized thin film. Generalizing
to the case of a spin current is straightforward, so long as we
assume the electrons in the current do not interact with one
another. This will allow us to address detailed questions about how
the spins in the chain reverse their magnetization when subjected to
a spin current. These and other questions will be the subject of
future investigation.

\begin{acknowledgments}

We wish to thank J. Hirsch for early correspondence concerning the
solution of the Schrodinger equation as presented in Ref.
\onlinecite{hirsch94}. W.K. acknowledges A. Mann and M. Revzen for
helpful discussions. This work was supported in part by the Natural
Sciences and Engineering Research Council of Canada (NSERC), by
ICORE (Alberta), and by the Canadian Institute for Advanced Research
(CIAR). F.M. is appreciative of the hospitality of the Department of
Condensed Matter Physics at the University of Geneva.
\end{acknowledgments}

\bibliographystyle{prb}

\begin{thebibliography}{1}

\bibitem{drescher} M. Drescher, M. Hentschel, R. Kienberger, M. Uiberacker, V. Yakovlev,
A. Scrinzi, Th. Westerwalbesloh, U. Kleineberg, U. Heinzmann and F. Krausz,
Nature {\bf 419}, 803 (2002)

\bibitem{baltuka} A. Baltuka, Th. Udem, M. Uiberacker, M. Hentschel, E. Goulielmakis,
Ch. Gohle, R. Holzwarth, V. S. Yakovlev, A. Scrinzi, T. W. Hansch and F. Krausz,
Nature {\bf 421}, 611 (2003).

\bibitem{hegman03} F.A. Hegmann, Physics in Canada {\bf 59}, 127 (2003).

\bibitem{freeman} M.R. Freeman and W.K. Hiebert, in {\it Spin Dynamics in Confined Magnetic
Structures I}, edited by B. Hillebrands and K. Ounadjela (Springer, Berlin, 2002).

\bibitem{choi01} B.C. Choi, M. Belov, W.K. Hiebert, G.E. Ballentine, and M.R.
Freeman, Phys. Rev. Lett. {\bf 86}, 728 (2001).

\bibitem{rikvold02} See, for example, P.A. Rikvold, G. Brown, S.J. Mitchell, and M.A.
Novotny, in {\it Nanostructured Magnetic Materials and their
Applications}, edited by D. Shi, B. Aktas, L. Pust, and F. Mikailov,
Springer Lecture Notes in Physics, Vol. 593 (Springer, Berlin,
2002), p. 164. See also cond-mat/0110103.

\bibitem{trugman03} L.-C. Ku and S. A. Trugman, cond-mat/0310226, unpublished.

\bibitem{berger96}
L. Berger, Phys. Rev. B{\bf 54}, 9353, (1996).

\bibitem{slonczewski96}
J.C. Slonczewski, J. Magn. Magn. {\bf 159} L1, (1996).

\bibitem{slonczewski99}
J.C. Slonczewski, J. Magn. Magn. {\bf 195} L261, (1999).

\bibitem{berger01} L. Berger, J. Appl. Phys. {bf 89} 5521 (2001).

\bibitem{llg1} J.Z. Sun, \prb {\bf 62}, 570 (2000).

\bibitem{llg2} J. Miltat, G. Albuquerque, and A. Thiaville,
in {\it Spin Dynamics in Confined Magnetic
Structures I}, edited by B. Hillebrands and K. Ounadjela
(Springer, Berlin, 2002).

\bibitem{kim05} W. Kim and F. Marsiglio, Europhys. Lett. {\bf 69} 595 (2005).

\bibitem{kim04} W. Kim and F. Marsiglio, \prb {\bf 69} 212406 (2004).

\bibitem{stiles} M.D. Stiles and A. Zangwill, \prb {\bf 66}, 014407 (2002).

\bibitem{shankar80} See, for example, R. Shankar, {\it Principles of
Quantum Mechanics}, (Plenum Press, New York, 1980).

\bibitem{optics} See, for example, G.P. Agrawal, {\it Nonlinear fiber optics}
(Academic Press, New York 2001).
We thank F. Hegmann for pointing this out to us.

\bibitem{lipkin73} H.J. Lipkin, {\it Quantum Mechanics: New Approaches
to Selected Topics} (North-Holland, Amsterdam 1973).

\bibitem{hirsch94} J.E. Hirsch, Phys. Rev. B{\bf 50}, 3165 (1994).

\bibitem{kim06} W. Kim, L. Covaci, and F. Marsiglio, 
to appear in \prb (cond-mat/0601323).

\bibitem{dunlap} D.H. Dunlap, H.-L. Wu, and P. Phillips, \prl {\bf 65},
88 (1990).

\bibitem{wu1} H.-L. Wu, and P. Phillips, \prl {\bf 66}, 1366 (1991).

\bibitem{wu} H.-L. Wu, W. Goff, and P. Phillips, \prb {\bf 45},
1623 (1992).

\bibitem{giri} D. Giri, P.K. Datta, and K. Kundu, \prb {\bf 48}, 14113
(1993).

\bibitem{gilmore} For an excellent discussion in the continuum limit,
see R. Gilmore, {\it Elementary Quantum Mechanics in One Dimension}
(The Johns Hopkins University Press, Baltimore, 2004).

\bibitem{burrow} B.L. Burrow and K.W. Sulton, \prb {\bf 51}, 5732 (1995).

\bibitem{avishai} Y. Avishai and Y. Tokura, \prl {\bf 87}, 197203 (2001).

\end{thebibliography}

\end{document}